\documentclass[useAMS,usenatbib]{mn2e}
\usepackage{aas_macros}
\usepackage{amsmath}
\usepackage{amssymb}
\usepackage{graphicx}

\begin{document}

\title[PISNe via collision runaway]{Pair-Instability Supernovae via Collision Runaway in Young Dense Star Clusters}
\author[T. Pan,  A. Loeb, and D. Kasen]{Tony Pan$^1$, Abraham Loeb$^1$, and Daniel Kasen$^{2,3}$\\
$^1$Harvard-Smithsonian Center for Astrophysics, 60 Garden Street, Cambridge, MA 02138, USA\\
$^2$Departments of Physics and Astronomy, 366 LeConte Hall, University of California, Berkeley, CA 94720, USA \\
$^3$Nuclear Science Division, Lawrence Berkeley National Laboratory, 1 Cyclotron Road, Berkeley, CA, 94708, USA
}

\pagerange{\pageref{firstpage}--\pageref{lastpage}} \pubyear{2011}

\maketitle

\label{firstpage}
\begin{abstract}
Stars with helium cores between $\sim$64 and 133 $M_{\odot}$ are theoretically predicted to die as pair-instability supernovae.  This requires very massive progenitors, which are theoretically prohibited for Pop II/I stars within the Galactic stellar mass limit due to mass loss via line-driven winds.  However, the runaway collision of stars in a dense, young star cluster could create a merged star with sufficient mass to end its life as a pair-instability supernova, even with enhanced mass loss at non-zero metallicity.  We show that the predicted rate from this mechanism is consistent with the inferred volumetric rate of roughly $\sim 2\times 10^{-9}$ Mpc$^{-3}$ yr$^{-1}$ of the two observed pair-instability supernovae, SN 2007bi and PTF 10nmn, neither of which have metal-free host galaxies.  Contrary to prior literature, only pair-instability supernovae at low redshifts $z<2$ will be observable with the \emph{Large Synoptic Survey Telescope} (LSST).  We estimate the telescope will observe $\sim 10^2$ such events per year that originate from the collisional runaway mergers in clusters. 
\end{abstract}
\label{lastpage}


\section{Introduction}

Pair-instability supernovae (PISNe) are thought to occur for stars with helium cores between $\sim$64 and 133 $M_{\odot}$ \citep{Heger2002}.  At zero metallicity, this corresponds to initial stellar masses between $\sim 140$ and 260 $M_{\odot}$.  These enormous stellar masses may have been reached by Pop III stars, predicted to have a top-heavy mass distribution \citep{Bromm2004}.  However, at lower redshifts, as the universe was enriched, Pop III stars ceased to form once the local metallicity exceeded a critical threshold $Z_{crit} \sim 10^{-3} Z_{\odot}$ \citep{Bromm2003}.  Since it is almost impossible to raise the intergalactic medium metallicity in a homogeneous way \citep{Furlanetto2003, Scannapieco2003}, pristine metal-free stars will still be formed past the end of the reionization epoch $z \lesssim 6$ \citep{Trenti2009}, conceivably all the way down to $z=2.5$ \citep{Tornatore2007}.  Observations have confirmed the existence of extremely metal-poor star formation at moderate redshifts of $z=3.357$ and $z=5.563$ \citep{Fosbury2003, Raiter2010}.  The detectability of PISNe from Pop III stars at these moderate redshifts was investigated by \citet{Scannapieco2005a}.

Outside these surviving pristine regions, there is a wide range of observations that support an upper limit to stellar mass at $\sim 150M_{\odot}$ in our Galactic neighborhood\citep{Figer2005, Weidner2010}, preventing the formation of PISNe from Pop II/I stars (but see \citet{Crowther2010} for stars determined to be above $150M_{\odot}$ in the R136 cluster.)  Nevertheless, even if very massive stars can form in metal rich regions, these radiatively supported stars are loosely bound and have strong winds driven mainly by radiation pressure through spectral lines, scaling as $\dot{M}\propto Z^{0.5\sim 0.7}$ \citep{Vink2001, Kudritzki2002}.  So even Pop II/I stars with initial masses between $\sim 140$ and 260 $M_{\odot}$ will suffer copious mass loss during both the hydrogen and helium burning stages, and may not end their lives with enough mass remaining to die as PISNe; this prediction could be contested, as there are still large uncertainties in mass loss models from hot massive stars \citep{Puls2008}.  Nevertheless, due to mass loss the possibility of PISNe is usually not considered for solar composition stars, even though the pair instability arises irrespective of the progenitor's metallicity.

Regardless, PISNe have very likely already been observed at low redshifts, most convincingly in the case of the very luminous and long duration event SN 2007bi \citep{Gal-Yam2009}.  More recently, the Palomar Transient Factory observed a new presumed PISN, PTF 10nmn (Gal-Yam 2011, submitted; Yaron et al., in preparation), and another PISN candidate was reported by the Pan-STARRS1 survey, PS1-11ap (Kotak et al., in preparation.)  Although it may be possible to explain bright events like SN 2007bi with alternative models (e.g. \citet{Woosley2007}, \citet{Moriya2010}, \citet{Kasen2010}) on the whole the observations seem to favor a scenario in which a large total mass and radioactive mass were ejected, as in a PISN explosion.  The observations  therefore suggest that very massive stars above the Galactic limit are formed in the local universe.  The metallicities of the host galaxies of both supernovae are low but well above the maximum metallicity required to form Pop III stars \citep{Young2010}.  Either pockets of pristine gas survived in the dwarf host galaxy of SN 2007bi and PTF 10nmn, or the initial mass function (IMF) of Pop II/I stars merely steepens at the very high end (instead of a hard upper limit), or there are other exotic ways to form a very massive star.

In theory, mergers of stars can form massive SN progenitors at any metallicity and circumvent the upper mass limit for Pop II/I stars at $\sim 150M_{\odot}$.  The most likely environment for such mergers is a dense, young star cluster undergoing core-collapse, in which a runaway collision product can become massive enough to die as an ultra-luminous supernova.  \citet{PortegiesZwart2007} first investigated this scenario for a collapsar, and \citet{Yungelson2008, Glebbeek2009, Vanbeveren2009} discussed the conditions under which the runaway collision product will end its life as a PISN.

In this paper, we calculate the number of the collision runaway merger products within dense young star clusters that lie in the PISN progenitor mass range, and show that the predicted event rate is roughly equal to the inferred rate of PISNe from the detection of SN 2007bi and PTF 10nmn in existing surveys, without requiring the supernova progenitor to be metal free.  We further investigate the observability and rate of these events in the low redshift universe with the \emph{Large Synoptic Survey Telescope} (LSST).

\section{Rates from Runaway Collisions}

An appreciable fraction of stars are born in clusters; \citet{Bastian2008} found the fraction of mass that forms in clusters $>100M_{\odot}$ out of the total star formation rate to be $\Gamma \sim 8\pm 3\%$.  As soon as the cluster forms, the massive stars start to sink to the cluster center due to dynamic friction, driving the cluster into a state of core collapse on a timescale of $t_{cc} \sim 0.2 t_{rh}$, where $t_{rh}$ is the relaxation time:
\begin{eqnarray}
\nonumber  t_{rh}	&\approx& 2 {\rm Myr} \left(\frac{r}{1 {\rm pc}}\right)^{\frac{3}{2}}\left(\frac{m}{1 M_{\odot}}\right)^{-\frac{1}{2}}\frac{N}{\log\lambda}\\
						&\approx& 200 {\rm Myr} \left(\frac{r}{1 {\rm pc}}\right)^{\frac{3}{2}}\left(\frac{m}{10^6 M_{\odot}}\right)^{\frac{1}{2}}\frac{\langle m \rangle}{M_{\odot}}.
\label{RelaxationTimeEq}
\end{eqnarray}
Here $m$ is the cluster mass, $r$ is its half mass radius, $N$ is the number of stars, $\langle m \rangle = N/m \approx 0.5 M_{\odot}$ is the average stellar mass, and $\log\lambda \approx \log(0.1 N) \sim 10$ \citep{PortegiesZwart2010}.  In sufficiently compact clusters, the formation of a dense central subsystem of massive stars may lead to a collision runaway, where multiple stellar mergers result in the formation of an unusually massive object \citep{Gurkan2004, Freitag2006}.  This prescription is often invoked to form intermediate-mass black holes via the photodisintegration instability that collapses a super-massive star directly into a black hole.

For a successful collision runaway to occur, the star cluster must experience core collapse before the most massive stars explode as a SN ($\sim$3 Myr).  For compact clusters ($t_{rh} \lesssim 100$ Myr), basically all massive stars sink to the cluster core during the runaway, and the final merged object's mass scales with the cluster mass, $m_{r} \approx 8 \times 10^{-4} m \log\lambda$ \citep{PortegiesZwart2002}.  For clusters with longer relaxation times, only a portion of massive stars sink to the core in time and the merged object's mass scales as $m t_{rh}^{-1/2}$ \citep{McMillan2004}.  A fitting formula for combining these scalings is given by \citet{PortegiesZwart2006}, calibrated by N-body simulations for Salpeter-like mass functions:
\begin{equation}
m_r \sim 0.01m(1+\frac{t_{rh}}{100 {\rm Myr}})^{-\frac{1}{2}}.
\label{RunawayMassEq}
\end{equation}
To get statistics on the final runaway mass $m_r$ from equations (\ref{RelaxationTimeEq}) \& (\ref{RunawayMassEq}), we need to specify the number distribution of clusters as a function of their mass $m$ and radius $r$.  

The functional form of the cluster initial mass function is well represented by a \citet{Schechter1976} distribution,
\begin{equation}
\Phi(m)=\frac{dN}{dm}=Am^{-\beta}e^{-m/m_{*}},
\end{equation}
where observationally $\beta \sim 2$ \citep{Zhang1999, McCrady2007}.  For Milky Way-type spiral galaxies the break mass $m_{*} \approx 2\times 10^5 M_{\odot}$ \citep{Gieles2006, Larsen2009}, whereas for interacting galaxies and luminous IR galaxies $m_{*}\gtrsim 10^6 M_{\odot}$ \citep{Bastian2008}.  Our results are not sensitive to the choice of $m_{*}$.

\begin{figure}
\centering
\includegraphics[width=1\columnwidth]{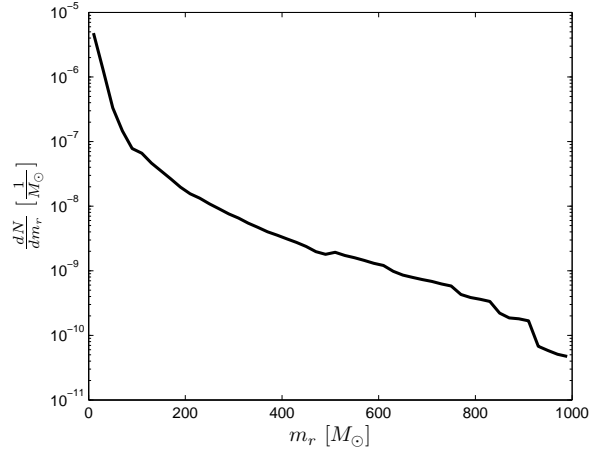}
\caption{Differential number distribution of the final runaway mass formed, \emph{per $1 M_{\odot}$ of stellar mass formed in all clusters}.  The calculated distribution is not perfectly smooth owing to the finite number of samples in the observed radius distribution.}
\label{PISNRateCollisionRunawayMassDistribution}
\end{figure}

Several studies have discussed the lack of any clear correlation between the size of a cluster and its mass or luminosity \citep{Larsen2004, Scheepmaker2007}.  Lacking a functional distribution of cluster radii, we use the empirical distribution of radii for each cluster mass bin, for observed clusters younger than $5$ Myr, compiled in Tables 2-4 of \citet{PortegiesZwart2010}.  The restriction on cluster age is important, as clusters expand considerably during the first 10 Myr of their evolution.  Note that this empirical construction underestimates the number of super-massive collision runaway objects ($>10^3 M_{\odot}$), as there happens to be no observed $>10^6 M_{\odot}$ clusters younger than $5$ Myr in the current sample, but this does not drastically affect our PISN rate estimates.  With the joint number distribution of clusters as a function of their mass and radii, we can find the number distribution of the final mass of the runaway collision merged object, see Figure \ref{PISNRateCollisionRunawayMassDistribution}.

\begin{figure}
\centering
\includegraphics[width=1\columnwidth]{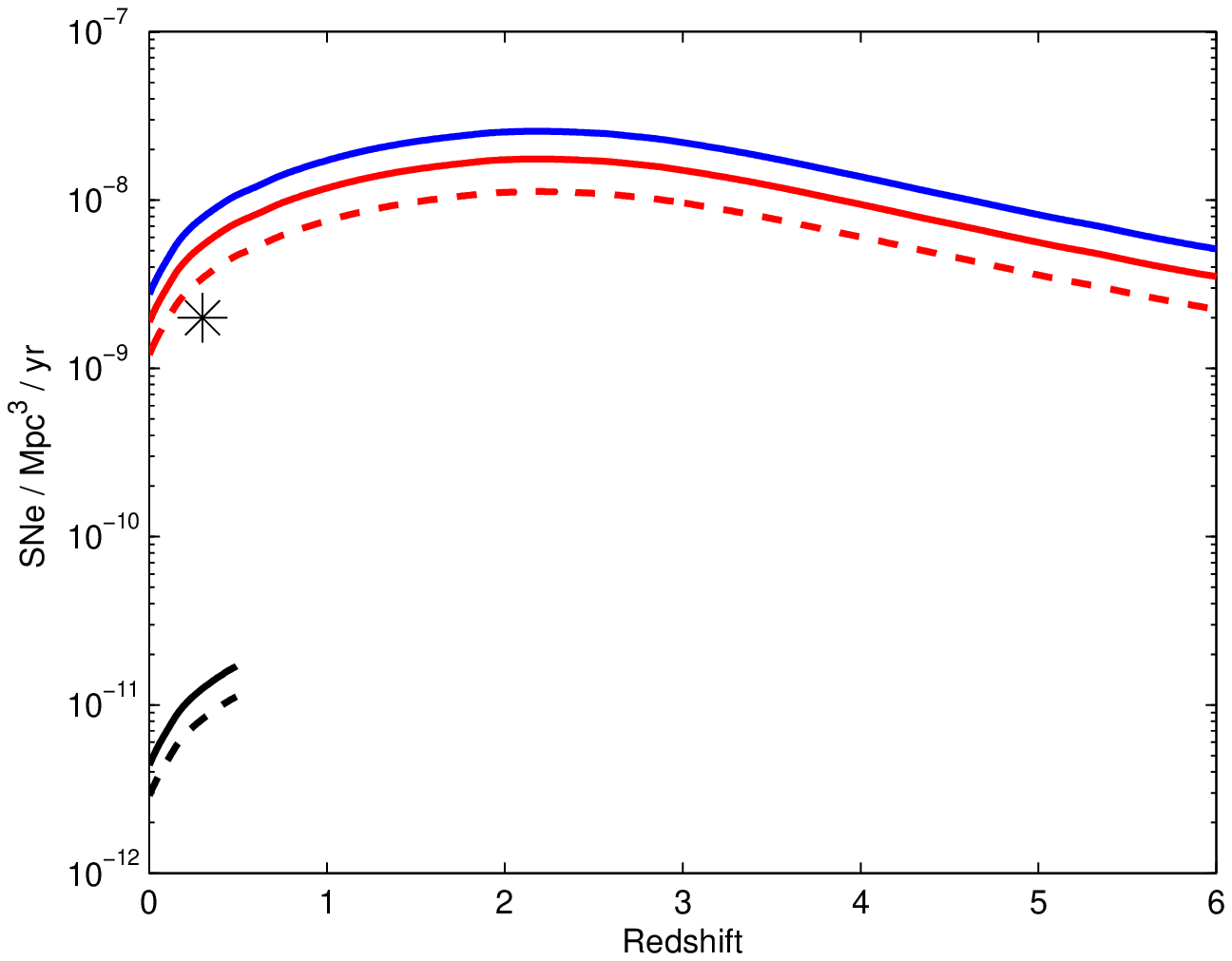}
\caption{ Predicted rate of PISN events per comoving Mpc$^3$ per year.  The pair-instability SNe progenitor mass range is a major uncertainty for non-pristine stars.  Here we use 140-260 $M_{\odot}$ (blue line) for metal-free Pop III stars from the models of \citet{Heger2002}, whereas the mass range of $\sim$250-800 $M_{\odot}$  (red line) is taken from \citet{Yungelson2008}, who account for increased mass loss at solar metallicity.  The environments of low redshift PISNe will likely lie in between these two cases.  A stronger mass loss scenario is presented by \citet{Belkus2007}, who found that PISNe progenitors can only be created at metallicities below 0.02 $Z_{\odot}$, with a mass range $\sim$300-1000 $M_{\odot}$ (black line); as the fraction of matter in $Z<0.02$ $Z_{\odot}$ galaxies is not well constrained past low redshifts, we do not plot this rate past $z=0.5$.  The strongest mass loss scenarios presented by \citet{Glebbeek2009} and \citet{Vanbeveren2009} predict a PISNe rate of practically zero, so we do not plot it here.  The dashed lines are the PISNe rates of \citet{Yungelson2008} and \citet{Belkus2007} adjusted for mass loss from stellar collisions.  The black star shows the inferred PISN rate from current surveys.}
\label{PISNRateCollisionRunawayVolumetricRate}
\end{figure}

\begin{figure}
\centering
\includegraphics[width=1\columnwidth]{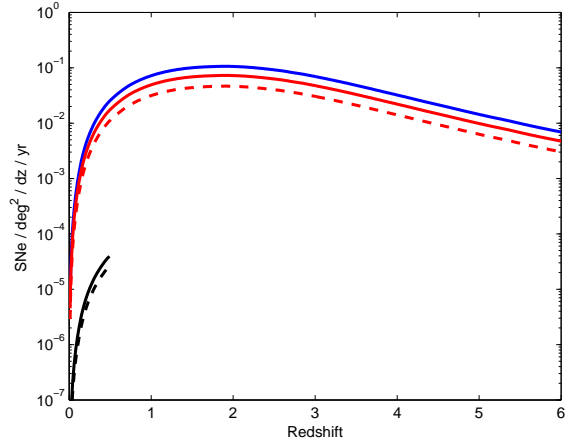}
\caption{Number of new PISNe per deg$^2$ per unit redshift per year, for the same models as Figure \ref{PISNRateCollisionRunawayVolumetricRate}.  Note that LSST is expected to cover over 20,000 deg$^2$ of sky.}
\label{PISNRateCollisionRunawaySqDegreeRate}
\end{figure}

However, as we have not taken mass loss into account in our estimate of the final runaway mass in equation (\ref{RunawayMassEq}), we artificially inflate the mass range for PISN progenitors required at the end of the last merger event, to compensate for the mass lost during the collision runaway merger sequence.  For zero-metallicity Pop III stars, mass loss via line-driven winds should be negligible, and \citet{Heger2002} found the progenitor mass range to be 140-260 $M_{\odot}$; this should set the upper limit of the PISNe rate from runaway collision products.  

As for Pop II/I PISNe progenitors, we caution that mass loss via
stellar winds for massive stars $M>100M_{\odot}$ is not well
understood.  In fact, the observations of PISNe at low redshifts
\citep{Gal-Yam2009}, and of Type IIn SNe whose progenitors are found
to sometimes retain their hydrogen envelopes until shortly before
their explosion \citep{Smith2011}, suggest that most commonly-used
stellar mass loss models are inaccurate for very massive stars, and
likely overestimate the total mass loss, as the models do not allow
such SNe to exist at the measured metallicities.  Therefore, to
account for this uncertainty, we present here various PISN progenitor
mass range scenarios described in literature, dependent on the assumed
metallicity and mass loss prescription.

\citet{Yungelson2008} studied the evolution and fate of super-massive
stars with solar metallicity from the zero-age main sequence using
detailed stellar structure models.  However, instead of extrapolating
commonly used mass loss models, e.g. \citet{deJager1988, Vink2001,
Kudritzki2002}, \citet{Yungelson2008} used an ad-hoc mass-loss
prescription consistent with existing models in their relevant regimes
and more consistent with the observed Hertzsprung-Russell diagram
location and mass loss ranges found for young massive stars in
clusters in the Milky Way and the Magellanic Clouds.  Notably, their
time-averaged Wolf-Rayet (WR) mass loss rate $\dot{M}_{WR}$ hardly
exceeds $10^{-4}$ $M_{\odot}$ yr$^{-1}$, which better fits
observations of hydrogen-rich WR stars that account for iron-line
blanketing and clumping in determining $\dot{M}_{WR}$
\citep{Hamann2006}, and also agrees well with $\dot{M}_{WR}$ estimates based on radio observations \citep{Cappa2004}. 
On the contrary, the extrapolation of Wolf-Rayet mass loss formulas to high stellar masses given by \citet{Langer1989},
\citet{Nugis2000}, \citet{Nelemans2001}, and \citet{deDonder2003}
overestimate the mass loss rates compared with these observations.
Therefore, we use the results of \citet{Yungelson2008} as our fiducial
model.  They allow the creation of PISNe progenitors at $Z\sim
Z_{\odot}$ in the initial mass range of $\sim$250-800 $M_{\odot}$;
however, they do not account for the mass loss from stellar
collisions.

Alternatively, by extrapolating theoretical mass-loss rates for
radiation-driven wind, \citet{Belkus2007} found that when the
metallicity $Z$ is between 0.001 and 0.02 $Z_{\odot}$, one may expect
PISN candidates for stars with masses from $\sim$300-1000 $M_{\odot}$;
however, at $Z>0.02$ $Z_{\odot}$ no PISNe are expected.  Using the
observed galaxy luminosity-metallicity relationship
\citep{Kirby2008,Guseva2009} and the galaxy luminosity function at low
redshifts $z\sim 0.1$ \citep{Blanton2003}, we find that $\sim 0.3\%$
of stellar mass is formed in $Z \lesssim 0.02$ $Z_{\odot}$ galaxies,
and fold this factor into the predicted PISNe rate for this scenario.

In addition, \citet{Glebbeek2009} follows the evolution of the
collision product for a few merger sequences for a $m \sim 5\times
10^5 M_{\odot}$ cluster, including mass loss along the course of the
collision sequence by using the prescription of \citet{Vink2001}, and
found that above $Z=0.001$ $Z_{\odot}$, the collision runaway product
cannot die with sufficient mass to undergo a PISN.  The main sequence
stellar wind mass loss rate between this work and \citet{Yungelson2008} are
similar, however, \citet{Glebbeek2009} also calculates the mass loss
from stellar collisions to be roughly $\sim 20\%$ of the total merger
product mass before mass loss.  Nevertheless, the main source of
discrepancy between their conclusions is due to their very different
Wolf-Rayet mass loss rates.  \citet{Glebbeek2009} implements a strong
Wolf-Rayet mass loss rate from \citet{Nugis2000} (up to $3.6\times
10^{-3}$ $M_{\odot}$ yr$^{-1}$ at $Z=0.02$), bringing the collision
product down to only $m_r \sim 10 M_{\odot}$ by the end of core
helium-burning.  Using a comparable mass loss rate,
\citet{Vanbeveren2009} reaches the same conclusion that PISNe cannot
occur above $Z=0.001$ $Z_{\odot}$.  Note that with these mass loss
rates, essentially no star in the low redshift universe below
$M\sim$1000 $M_{\odot}$ will end their lives as a PISN, irrespective
of the collision runaway mechanism.

To account for the $\sim 20\%$ mass loss due to unbound ejecta from
the stellar collision, we can further increase the required PISN
progenitor mass range.  The new adjusted mass range for PISN
progenitors would be $\sim$313-1000 $M_{\odot}$ in the
\citet{Yungelson2008} scenario, and $\sim$375-1250 $M_{\odot}$ in the
\citet{Belkus2007} scenario.

Combining the above, we can estimate the number of collision runaway
products that have a final mass $m_r$ in the various PISN progenitor
mass range scenarios.  Using the global comoving star formation rate
from \citet{Reddy2009}, we estimate the PISN rate as a function of
redshift in Figures \ref{PISNRateCollisionRunawayVolumetricRate} and
\ref{PISNRateCollisionRunawaySqDegreeRate}.  If the collision runaway
mechanism is indeed responsible for creating PISNe progenitors at Pop
II/I environments in the local universe, we find that only the mass
loss prescription described by \citet{Yungelson2008} fits the current
rate of PISNe inferred from observation.

\section{Observability with LSST}

The \emph{Large Synoptic Survey Telescope} is a planned wide-field survey telescope that should begin operations at the end of this decade.  It has a very wide field of view of $9.6$ deg$^2$, and 6 bands: $u, g, r, i, z$, and $y$, covering 320-1080 nm.  For the most sensitive bands $g, r$, and $i$, a single visit will reach $M_{AB}=25.0$, 24.7, and 24.0 (5$\sigma$) sensitivity, respectively.  These bands will be visited 10, 23, and 23 times every year during the 10 years of operation, reaching a coadded depth of $M_{AB}=$26.3, 26.4, and 25.7 per year by stacking multiple images.  Note that for objects much dimmer than $\sim 22$ mag/arcsec$^2$, or $\sim 25.5$ mag/pixel for LSST, the signal will be dominated by the sky background (e.g. airglow and zodiacal light), so in this regime the limiting signal flux needed to reach a fixed signal-to-noise ratio is inversely proportional to the square root of the integration time.

We use simulated PISN light curves and spectra from \citet{Kasen2011}, who improved radiative transfer calculations by using a multi-wavelength Monte Carlo code which includes detailed line opacities.  In particular, we use models He130, He100, and He80, which represent pair-instability explosions of non-rotating bare helium cores with masses 130, 100, and 80 $M_{\odot}$, respectively, as non-pristine massive stars formed via runaway collision will likely lose most of their hydrogen envelope by the end of their life.  The brightest helium core model He130 peaks at around $M_{AB} \sim -22$ in the rest frame r band, and stays above $M_{AB}=-21$ for half a year, and above $M_{AB}=-20$ for almost one year.  Such an event in the local universe will be easily detectable; however, the rates for PISNe from both Pop III and Pop II/I progenitors is predicted to be very low at $z\sim 0$.  These rates increase at higher redshifts, but since the higher wavelength $z, y$ LSST bands are much less sensitive, the best strategy to find PISNe is to continue using the $g, r$, and $i$ bands and observe at the rest frame UV and optical luminosity of the supernovae.

\begin{figure}
\centering
\includegraphics[width=1\columnwidth]{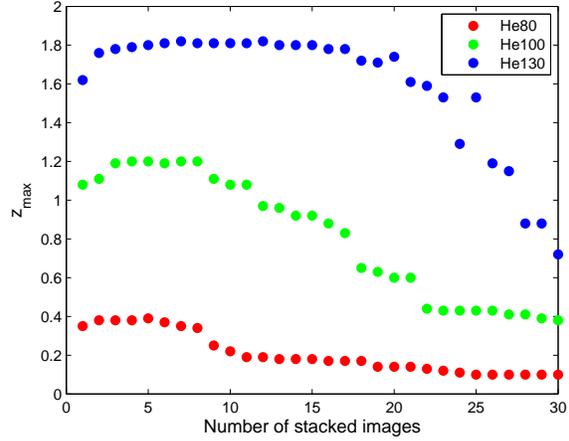}
\caption{Maximum redshift observable by LSST, as a function of the number of stacked images, for various PISN progenitor models.  Here we use the co-added $r$ band 5$\sigma$ sensitivities, for which LSST will visit the same location 23 times every year, or once every $\sim 16$ days on average.  We consider a PISN at a certain redshift as observable if it stays brighter than the limiting co-added depth for a duration longer than the time it takes to observe that number of images.  $z_{max}$ eventually drops with increasing image count, as the PISN flux falls off but the sky background remains, reducing the integrated signal-to-noise.  The brightest PISN will be observable with LSST out to a redshift of $\sim 1.8$.}
\label{LSST_max_redshift}
\end{figure}

Using the coadded depth sensitivities, we find that using the $r$ band is optimal for the helium core PISN models, and that we can observe the  brightest He130 model out to a redshift of $z\sim 1.8$ by stacking $\sim 10$ images (see Figure \ref{LSST_max_redshift}).  Below $z<2$, the PISN is visible in the $r$ band for over 1 year in the observer frame; however, at $z\geq 2$, the supernova will be too dim in the rest frame UV wavelengths being effectively probed, even though the $(1+z)$ time dilation allows more stacked images.  Even if one combines data from the $g, r$, and $i$ bands over one year, and reaches a coadded depth of $M_{AB}\sim 27$, the supernova will still be too dim to be observable beyond $z=3$.  Alternative PISN models where the progenitors are red supergiants which retain their hydrogen envelopes have a longer plateau in their light curves, and thus stay visible slightly longer than the helium core models.  However, the conclusions are similar - in terms of instrument capability, redshifts $z<2$ are most suitable for detecting PISNe in the normal operation mode of LSST.  The smaller He100 model is only visible out to $z\sim 1.2$, while even smaller progenitors are too dim to be seen beyond $z<0.4$.  Combined with Figure \ref{PISNRateCollisionRunawaySqDegreeRate}, we estimate that LSST will see on the order of $\sim 10^2$ new PISNe per year that originated from the final collision runaway object in young, dense clusters.

These conclusions differ from those of \citet{Trenti2009} as well as the LSST Science Book \citep{LSSTScienceCollaborations2009}, which concluded that PISN at $z\sim 4$ will be within the capability of LSST.  The difference arises because \citet{Trenti2009} approximated the PISN with a blackbody spectrum with $T_{eff}=1.5\times 10^4$K, which overestimates the rest frame UV flux compared to the spectrum obtained by the radiation hydrodynamics simulations of \citet{Kasen2011}.  Also, in the LSST Science Book, when calculating that hundreds of $z=2-4$ PISNe will be detected by LSST (Chapter 11.14), the authors used $z$ and $y$ band sensitivies of $\sim 26.2$.  This is unrealistic as $M_{AB} \sim 26.2$ can only be reached in the $z$ band by stacking all images over the entire 10 year lifetime of the survey, but no PISN will stay bright enough that long even with time dilation; the $y$ band is even less sensitive.  Our findings suggest that, to find PISNe at $z>2$, an instrument with better infrared capabilities such as the \emph{James Webb Space Telescope}\footnote{http://www.jwst.nasa.gov/} is required.

Although stacking multiple images averages out the time variation in the supernova light curve, LSST also allows a secondary survey over a smaller area of sky, going substantially deeper in a single epoch.  However, due to the steep luminosity function of PISNe, we will preferentially see only the massive PISN events beyond the local universe, so narrow, deep exposures by LSST are more useful for improving light curve coverage, instead of supernova discovery.

\section{Discussion}

Runaway collisions were explored most seriously in massive, dense clusters, so equation (\ref{RunawayMassEq}) may not be accurate for $m<10^4 M_{\odot}$.  However, only more massive clusters can make runaway masses $m_r$ in the PISN progenitor mass range, so this does not affect the predicted PISN rate.  In addition, initial mass segregation of stars within young clusters observed by \citet{Degrijs2002} and \citet{Stolte2006} will shorten the time to runaway collisions and increase $m_r$, but we do not take this into account.

For $z\lesssim 6$, the rate and detectability of PISN from Pop III progenitors born in surviving pockets of metal-free gas was investigated by \citet{Scannapieco2005a}.   To model the PISN light curves, they used an implicit hydrodynamics code which only implements gray diffusive radiation transport; for spectra and colors they assumed a blackbody distribution.  Depending on the intergalactic medium metal enrichment history, their predicted rates span two orders of magnitude, with their lower end roughly equal to our collision runaway rates at $z=1-2$.  However, a PISN with a pristine host galaxy has yet to be observed.

A pilot search done using the \emph{Spitzer}/IRAC dark field found no candidates above the sensitivity limit of $M_{AB}(3.6\mu m)\sim 24$, placing an 95$\%$ confidence upper limit of 23 per deg$^2$ per year for $>1$$\mu$Jy sources with plateau timescales less than $400/(1+z)$ \citep{Frost2009}, which does not contradict the predicted rate of $<0.1$ PISN per deg$^2$ per year for our collision runaway model.

More recently, observers have discovered a class of ultra-luminous supernova, with luminosities  exceeding those of the brightest pair-instability events, and rates of order $\sim 10^{-8}$ Mpc$^{-3}$ yr$^{-1}$at $z\approx 0.3$.   These events do not appear be standard radioactively powered PISNe, as their luminosities are too high and their light curve durations too short \citep[e.g.,][]{Quimby2011, Chomiuk2011}.  Comparing the rate of those events to that of the two putative observed PISNe, Gal-Yam found that PISNe are roughly $\sim$5 times rarer than the \citet{Quimby2011} ultra-luminous supernovae (Gal-Yam 2011, Science, submitted).  This gives a PISN rate of $\sim 2\times 10^{-9}$ Mpc$^{-3}$ yr$^{-1}$ in the local universe, roughly consistent with the collision runaway rates found in Figure \ref{PISNRateCollisionRunawayVolumetricRate}.

If the collision runaway of massive stars in young, massive stellar clusters do give rise to PISNe at Pop II/I metallicities, we expect to see such a young, massive cluster at the same location, after the light of the supernova has faded away.  However, even a $10^5 M_{\odot}$ cluster only has an absolute magnitude of about -8.2 mag, so the PISN will have to occur close by ($z<0.05$) for its host cluster to be observed with current telescopes.  Also, these PISNe should follow the distribution of clusters, and appear in the luminous parts of their host galaxies, analogous to the position of long duration gamma-ray bursts.  Note that due to the steep distribution of collision runaway masses (see Figure \ref{PISNRateCollisionRunawayMassDistribution}), the rates of PISNe from collision runaways will still be higher in environments with low metallicities, as long as mass loss for massive stars is proportional to metallicity.

Alternatively, if mass loss models are wrong and the Galactic stellar mass limit is violated, we need not invoke stellar mergers to create the massive progenitors required for the observed non-pristine PISNe. \citet{Langer2007} found that hydrogen-rich PISNe could occur at metallicities as high as $Z_{\odot}/3$, resulting in a rate of about 1 PISN per $10^3$ SNe in the $z\approx 0$ universe.  For a more conservative metallicity threshold of $Z_{\odot}/10$, the rate would be about 1 PISN per $10^4$ SNe.  However, even the latter is a few times higher than the current inferred rate of PISNe.

\section{Conclusion}

We have shown that the runaway collision and merger of stars in a young, dense star cluster may form the massive progenitor of a pair-instability supernova at non-zero metallicity.  The volumetric rate of such events is a few times $10^{-9}$ Mpc$^{-3}$ yr$^{-1}$ in the local universe, roughly matching the inferred rate of pair-instability supernova events SN 2007bi and PTF 10nmn in ongoing surveys, both of which have a metal-poor but not metal-free host galaxy.  We expect that the primary survey of the Large Synoptic Survey Telescope would see $\sim 10^2$ such events per year.

\section*{Acknowledgments.}
We thank Charlie Conroy for helpful comments.  TP was supported by the Hertz Foundation.  This work was supported in part by NSF grant AST-0907890 and NASA grants NNX08AL43G and NNA09DB30A.  
DK is supported in part by the Director, Office of Energy
Research, Office of High Energy and Nuclear Physics, Divisions of
Nuclear Physics, of the U.S. Department of Energy under Contract No.
DE-AC02-05CH11231, and by an NSF Astronomy and Astrophysics Grant  NSF-AST-1109896.
This research has been supported by the DOE
SciDAC Program (DE-FC02-06ER41438).  We are grateful for computer time provided by ORNL through an INCITE award and by NERSC.

\bibliographystyle{mn2e}
\bibliography{references}

\end{document}